\documentclass[english,a4paper,11pt]{article}

\usepackage{amssymb,amsmath,amsfonts,amsthm,epsfig,pstricks,graphics,tikz}

\usepackage[paper=a4paper,left=30mm,right=30mm,top=25mm,bottom =25mm]{geometry}

\usepackage[width=0.85\textwidth ]{caption}

\usepackage{hyperref}
\title{\LARGE{\textbf{{Classical and quantum billiard inside the square with gravitational field}}}}

\author{Daniel Jaud\\
\small Gymnasium Holzkirchen\\
\small \href{mailto:Daniel.Jaud.PhD@gmail.com}{Daniel.Jaud.PhD@gmail.com}}
\date{}

\begin{document}
\maketitle
%\flushbottom

%\hrulefill\\ %Zeile mit Linie füllen

%\twocolumn

\begin{abstract}
In this work the classical motion and quantum behavior of a particle inside a square of length $L$ under the influence of a gravitational field is considered. This includes a study for the conditions on classical periodic orbits as well as classical probability densities and associated position expectation values and standard deviations. In the quantum world the appropriate wave functions and energy eigenvalues are derived concluding a comparison to the classical obtained probability densities and expectation values.
\end{abstract}


%\vspace*{0.7cm}
%\hrulefill\\ %Zeile mit Linie füllen
%\tableofcontents
%\hrulefill\\
%\newpage
%\hrulefill\\

\section{Introduction}

During the last decades dynamical systems, especially billiard systems, have widely been studied in various context such as the theory of classical periodic orbits \cite{Rozikov,Tabach} in polygonal systems, quantum versions and visualizations \cite{Circular_Well} or applications to number theory \cite{Jaud1}. Due to simplicity those billiard systems are often studied in a force free scenario where the particles motion is determined by its initial conditions as well as the law of reflection at the boundary. The next interesting systems are those involving a constant force along one direction which e.g. can be induced by a gravitational field. Recent research on this topic has e.g. classically been performed for the case of a parabolic boundary in \cite{Masa_Grav} or in the quantum version either analytically in the simplest case \cite{Gupta} or numerically in \cite{Grav_Bill_various}. 

In this research we investigate the classical and quantum behavior of a particle inside a square box which is known to be integrable.

The structure of the paper is as follows: In section 2 we consider the classical motion of a particle inside a square with gravitational field along the $y-$direction and derive the conditions for periodic orbits. In the following section we study the classical probability density for this setup and derive the expected position as well as the standard deviation of the particle including a study of limiting cases for small and large vertical energies. Section 4 will be focusing on the quantum mechanical description resulting in a derivation of the associated wave functions and quantized energy eigenvalues. The paper concludes in the quantum mechanical study of the probability density and the associated expectation values for the position, including a comparison to the classical results.

\section{Classical Square Study}
In this section we focus on particle trajectories subjected to a linear increasing potential (in our case gravity). By a suitable substitution of variables this problem can also be interpreted as a charged particle in a constant electric field along one direction. 
\begin{figure}[htb]
    \centering
    \includegraphics[scale=0.6]{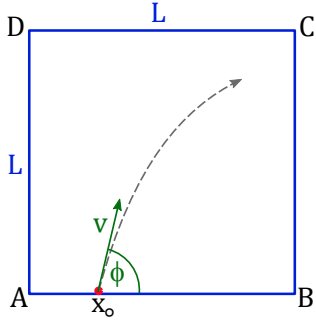}
    \caption{Particle inside square with initial conditions.}
    \label{fig:start_class}
\end{figure}
It turns out that periodic orbits are completely determined by the trajectory behavior along the boundaries. For a particle moving inside a square of side length $L$ under the influence of gravity along the $y-$axis the potential is given by
\begin{equation} \label{eq:square_potential}
    V(x,y)=\begin{cases}
    mgy & \mbox{for}~(x,y)\in [0;L]^2,\\
    \infty & \mbox{otherwise}.
    \end{cases}
\end{equation}
The motion of the particle is totally characterized by the initial starting position $x_0$, the absolute value of the velocity $v$ (or equivalently its total energy $E$) at $y=0$  and the angle $\varphi$ under which the particle ejects into the domain (see Figure \ref{fig:start_class}).

The equations of motion for time $t$ before the first scattering along the boundary are given by
\begin{align}
    x(t)&=v\cdot \cos(\varphi)\cdot t+x_0,\\
    y(t)&=-\frac{1}{2}g\cdot t^2+v\cdot \sin(\varphi)\cdot t.
\end{align}

It turns out that rewriting $y$ in terms of $x$ and replacing the velocity dependence by the associated total energy of the system $E=\frac{1}{2}mv^2$ is an advantage. As a result, one obtains the particle height in terms of its distance along the $x-$direction via
\begin{equation}
    y(x)=-\frac{mg}{4E\cdot \cos^2(\varphi)}\cdot (x-x_0)^2+\tan(\varphi)\cdot (x-x_0).
\end{equation}

\begin{figure}[htb]
    \centering
    \includegraphics[scale=0.56]{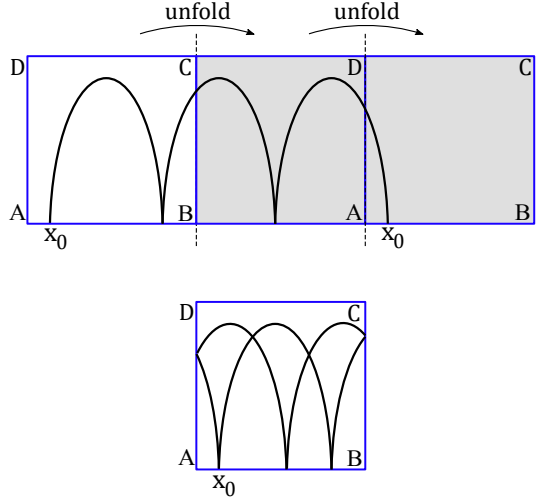}
    \caption{\textit{Above:} Periodic trajectory via unfolding procedure for $E\leq mgL$. \textit{Below:} Associated periodic trajectory in square.}
    \label{fig:folding1}
\end{figure}

Trajectories thus correspond to parabolas. Hitting the left or right boundary simply corresponds to inverting the $x-$direction of the trajectory. The particle motion can thus be alternatively described by unfolding the square \cite{Jaud1,Rozikov,Tabach} and consider the alternating parabola trajectories in the unfolded picture (see Figure \ref{fig:folding1}). 

For energies $E< \frac{mgL}{\sin^2(\varphi)}$, which simply means that the particle energy in the $y-$direction is not enough to encounter the upper boundary and thus only propagating in domains $0\leq y<L$, periodic orbits thus arise if the difference between two consecutive zeros of $y(x)$, i.e. $\Delta x(y=0)$, satisfies
\begin{equation}\label{eq:simple_cond}
    p\cdot \Delta x(0) =q\cdot 2L~~~~~p,q\in \mathbb{N}^+.
\end{equation}
The factor 2 represents the fact that the domain has to be unfolded twice to return to its initial orientation (see Figure \ref{fig:folding1}).

\begin{figure}
    \centering
    \includegraphics[scale=0.56]{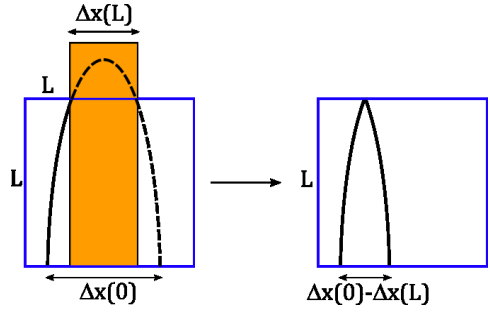}
    \caption{Reduced distance between two zeroes of $y(x)$ for energies larger or equal than $\frac{mgL}{\sin^2(\varphi)}$.}
    \label{fig:folding2}
\end{figure}

If the particles energy along the $y-$direction is large enough to obtain (in theory) heights ${y\geq L}$, the particle bounces at the upper boundary. Since in this case the $y-$component of the velocity is inverted ($v_y\rightarrow -v_y$)  the effective distance covered by two consecutive bounces at $y=0$ with one intermediate bounce at $y=L$  (see Figure \ref{fig:folding2}) is given by $\Delta x(0)-\Delta x(L)$, where

\begin{equation}
    \Delta x(y)=\frac{2E}{mg}\cdot \sin(2\varphi)\cdot \sqrt{1-\frac{mgy}{E\cdot \sin^2(\varphi)}}.
\end{equation}

Applying the unfolding procedure we can see that the particle describes periodic orbits iff the difference $\Delta x(0)-\Delta x(L)$ satisfies
\begin{equation}\label{eq:periodic_cond_final}
   [\Delta x(0)-\Delta x(L)]\cdot p=q\cdot 2L~~~~p,q\in \mathbb{N},
\end{equation}
i.e. periodic orbits in the square geometry with gravity are completely determined by the behavior of the particle along the boundaries at $y=0$ and $y=L$. Note that \eqref{eq:periodic_cond_final} also covers the first case displayed in Eq. \eqref{eq:simple_cond} with energies along the $y-$direction below $mgL$. In that scenario, it is implicitly understood that $\Delta x(L)=0$ if $E\cdot \sin^2(\varphi)\leq mgL$. For the particle trajectories not hitting a corner of the square, in which case the law of reflection is not defined, there are two further conditions - namely
\begin{equation}
 x_0+[\Delta x(0)-\Delta x(L)]\cdot (n+\delta) \neq m\cdot L,
\end{equation}
with $n,m\in \mathbb{N}$ and $\delta =0$ for hitting the corners $A$ or $B$ and $\delta=\frac{1}{2}$ for corners $C$ or $D$.

\section{Classical Probability Density and Expected Positions}
In this section we are first considering the probability density of a particle moving inside the square including gravity along the $y-$direction. The maximal theoretical height (neglecting the boundary) a particle might reach is given in terms of the total energy $E$, its mass $m$ and the initial angle $\varphi$ by
\begin{equation}
    h_{max}=\frac{E\cdot \sin^2(\varphi)}{mg}.
\end{equation}
Note that in the following simply abbreviate $h_{max}=h$. From here we start off by considering the classical probability density $\varrho(x,y)$ which has been discussed \cite{class_prob1,class_prob2} in various other physical systems e.g. free particles or the harmonic oscillator. Due to the fact that the motion in the $x-$direction is homogeneous, there is no dependence on $x$ in the probability density, i.e. $\varrho(x,y)=\varrho(y)$. Classically, the probability of finding the particle in the vicinity $dx\cdot dy$ is proportional to the differential time $dt$ of the particle in this vicinity, i.e.
\begin{equation}
    \varrho(y)dxdy\sim dt.
\end{equation}
The last equation is equivalent to
\begin{equation}
    \varrho(y)dxdy\sim \frac{dy}{\frac{dy}{dt}}=\frac{dy}{v_y(y)}.
\end{equation}
By conservation of energy, it is easy to verify that the velocity $v_y$ in $y-$direction depending on the energy $E$ and potential energy $V(y)=mgy$ of the particle at a height $y$ is proportional to
\begin{equation}
    v_y(y)\sim \sqrt{E\cdot \sin^2(\varphi)-mgy}.
\end{equation}
A direct calculation shows that the classical probability density of the particle written in terms of the maximal theoretical height $h$ is given by
\begin{equation}
    \varrho(y)= \frac{N}{\sqrt{h-y}}\cdot \Theta\left(h-y\right),
    \label{eq:classical_prob_density}
\end{equation}
where the Heaviside Theta-function is defined by
\begin{equation}
    \Theta(x)=\begin{cases} 0 &\mbox{for} ~x<0,\\
    1 &\mbox{for}~x\geq 0,
    \end{cases}
\end{equation}
and insures that the $y-$integration runs from zero to the maximal allowed height $h$ but not further than the boundary of the domain $L$. The correct normalization constant $N$ can, as usual, be obtained by the condition
\begin{equation}
    \int_0^L\int_0^{L}\varrho(y) dy dx=1,
\end{equation}
and takes the form
\begin{equation}\label{eq:class_prob_norm}
    N=\begin{cases}\frac{1}{2L\cdot \sqrt{h}} &\mbox{for} ~h<L,\\
    \frac{1}{2L\cdot \left(\sqrt{h}-\sqrt{h-L}\right)} &\mbox{for} ~h\geq L.
    \end{cases}
\end{equation}

\begin{figure}[htb]
    \centering
    \includegraphics[width=0.45\textwidth]{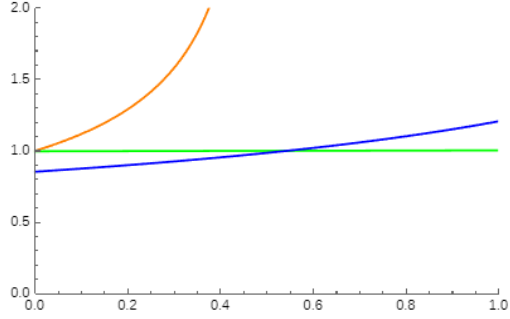}
    \caption{Classical probability densities along $y-$direction for $L=1$. \textit{orange}: $h=0,5$. \textit{blue}: $h=2$. \textit{Green}: Large energy limit where correspondingly $h$ becomes large.}
    \label{fig:class_prob_examples}
\end{figure}

Figure \ref{fig:class_prob_examples} displays some cases for the obtained probability density along the $y-$direction for\newline $L=1$ and various vertical energies. Obviously, the probability density approaches $1/L^2$ for large energies (\textit{green} curve) due to the reflection along the upper boundary and displays an asymptotic behavior (\textit{orange} curve) in the low energy limit since in this regime $y<L$.

Applying Eq. \eqref{eq:classical_prob_density} with associated normalization Eq.  \eqref{eq:class_prob_norm} it is now straight forward to calculate the classical expected particle positions $\langle \vec{r}\rangle_{cl}$ inside the box as a function of $h$ (i.e. corresponding energy) via
\begin{equation}
    \langle \vec{r}\rangle_{cl} =\int_0^L \int_0^L \vec{r}\cdot \varrho (y)dxdy.
\end{equation}
Note that the same expression can later on be applied for the quantum case study where the probability density is given by the absolute square of the wave function. Similarly, one can calculate the expectation values $\langle x^2\rangle_{cl}$ and $\langle y^2\rangle_{cl}$ which are needed for the calculation of the particles uncertainty (or standard deviation ${\Delta x_{cl}=\sqrt{\langle x^2\rangle_{cl}-\langle x\rangle_{cl}^2}}$ and analogous for $\Delta y_{cl}$. 

We will begin with a short discussion of the $x-$direction. Since in this direction there is no force acting on the particle, the probability density and associated expectation value calculations can easily be carried out, yielding
\begin{equation}\label{eq:class_x_expectation}
    \langle x\rangle_{cl} =\frac{L}{2},
\end{equation}
and
\begin{equation}
    \langle x^2\rangle_{cl}=\frac{L^2}{3}.
\end{equation}
As a direct result, the classical uncertainty of the particle along the $x-$direction is given by
\begin{equation}\label{eq:class_standard_x_deviation}
    \Delta x_{cl}=\frac{L}{2\sqrt{3}}.
\end{equation}

\begin{figure}[htb]
    \centering
    \includegraphics[scale=0.65]{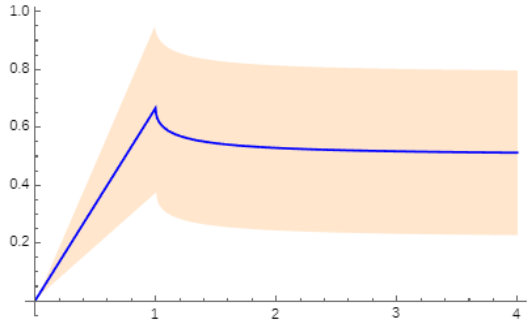}
    \caption{\textit{Blue:} Classical $\langle y\rangle_{cl}$ expectation value as a function of $h_{max}$ for $L=1$. \textit{Orange:} Associated domain $\langle y\rangle_{cl}\pm \Delta y_{cl}$.}
    \label{fig:classical_expectation}
\end{figure}

The results for the $y-$direction can also be performed, resulting in the expressions (again in terms the Heaviside function)
\begin{equation}\label{eq:y_exp_cl}
    \langle y\rangle_{cl}=\frac{2}{3}h-\frac{L}{3\cdot \left(\sqrt{\frac{h}{h-L}}-1\right)}\cdot \Theta(h-L),
\end{equation}
and
\begin{equation}
     \langle y^2\rangle_{cl}=\frac{8h^2}{15}-\frac{4hL+3L^2}{15\cdot \left(\sqrt{\frac{h}{h-L}}-1\right)}\cdot \Theta(h-L).
\end{equation}
Remarkably, the result for $\langle y\rangle_{cl}$ demonstrates that the maximal expected height of the particle is given by $\frac{2}{3}h$ and is accomplished if ${E\cdot \sin^2(\varphi)=mgL}$, i.e. if the vertical energy of the particle matches the potential energy at the upper boundary. For energies above $mgL$ the expectation value $\langle y\rangle_{cl}$ approaches $L/2$ which is the same as $\langle x\rangle_{cl}$ and is clear from the fact that in this limit the gravitational force is merely neglectable resulting basically in the movement of a free particle inside a square box. For the classical standard deviation this results in
\begin{equation}\label{eq:y_uncert_cl}
 \Delta y_{cl}=\frac{2\sqrt{15}}{15}\cdot h\cdot \sqrt{1+J(L,h)\cdot \Theta(h-L)},  
\end{equation}
where, in order to write the formula in a compact form, we defined the new function $J(L,h)$ corresponding to the correction terms for the standard deviation induced by the upper boundary compared to the system without the boundary by
\begin{equation}
J=\frac{L\left(8h\left(\sqrt{\frac{h}{h-L}}-1\right)+L\left(4-9\sqrt{\frac{h}{h-L}}\right)\right)}{4h^2\left(\sqrt{\frac{h}{h-L}}-1\right)^2}.
\end{equation}
Again, as for the expected particle position in $y-$direction in the large vertical energy limit, the standard deviation approaches the same value as in $x-$direction, i.e.
\begin{equation}
    \lim_{h\rightarrow \infty}\Delta y_{cl}=\Delta x_{cl}=\frac{L}{2\sqrt{3}}.
\end{equation}
A graphical representation for $\langle y\rangle_{cl}$ (\textit{blue} curve) and the associated uncertainty $\Delta y_{cl}$ (\textit{orange} domain) as a function of $h$ and the special case of a unit square with $L=1$ is displayed in Figure \ref{fig:classical_expectation}.

\section{Quantum Square Study}
The Schrödinger equation for a quantum particle of mass $m$ inside a square of length $L$ and potential given by Eq. \eqref{eq:square_potential} reads

\begin{equation} \label{eq:SGL_square}
    \left[-\frac{\hbar^2}{2m}\cdot \Delta_2+mgy\right]\cdot \Psi(x,y)=E\cdot \Psi(x,y),
\end{equation}

where $\Delta_2=\partial_x^2+\partial_y^2$ is the two dimensional Laplace-operator. Referring to the classical system, where the $x-$ and $y-$motion are independent from each other the wave function can be represented by the product of two separate functions for each variable, i.e. $\Psi(x,y)=X(x)\cdot Y(y)$. With this ansatz the Schrödinger equation basically falls apart into two independent equations for $X(x)$ and $Y(y)$ separately, namely
\begin{align} \label{eq:X_Y}
   -\frac{\hbar^2}{2m}\cdot  X''(x)&=E_x\cdot X(x),\\
   -\frac{\hbar^2}{2m}\cdot  Y''(y)+mgy\cdot Y(y)&=E_y\cdot Y(y), \label{eq:Y}
\end{align}
where $E_x$ and $E_y$ are the corresponding energy values in $x-$ and $y-$direction adding up to the total particle energy $E=E_x+E_y$.

It is obvious that allowed wave functions for the $x-$direction correspond to those of the infinite potential well
\begin{equation}
    X_n(x)=\sqrt{\frac{2}{L}}\cdot \sin\left(\frac{n\cdot \pi \cdot x}{L}\right),
\end{equation}
with $n\in \mathbb{N}^+$. The quantization in $x-$direction arises from the boundary conditions $X(0)=0=X(L)$. The associated quantized energy eigenvalues read
$$E_{x,n}=\frac{n^2\cdot \pi^2\cdot \hbar^2}{2mL^2}.$$

In order to find solutions to the $y-$direction wave function in Eq. \eqref{eq:Y} we first define new variables, namely
\begin{equation}
    R=\left(\frac{\hbar^2}{2m^2g}\right)^{\frac{1}{3}},
\end{equation}
as well as
\begin{equation}\label{eq:variables}
    z=\frac{y}{R}-\epsilon_{y}~~~~~\mbox{with}~~~~~\epsilon_{y}=\frac{2mR^2\cdot E_{y}}{\hbar^2}.
\end{equation}
In this new variables Eq. \eqref{eq:Y} becomes
\begin{equation}
    Y''(z)-z\cdot Y(z)=0.
\end{equation}

The last equation is the well-known Airy-equation. The solutions are the Airy-functions of first $\text{Ai}(z)$ and second kind $\text{Bi}(z)$. General properties and applications of the Airy-functions can e.g. be found in \cite{Airy_Book}. The general solution for the wave function in $y-$direction is thus given by
\begin{equation}
    Y(y)=c_1\cdot \text{Ai}\left(\frac{y}{R}-\epsilon_{y}\right)+c_2\cdot \text{Bi}\left(\frac{y}{R}-\epsilon_{y}\right).
\end{equation}
The boundary conditions at $y=0$ and $y=L$ on $Y(y)$ imply that
\begin{align}\label{eq:boundary_sol1}
    0&=c_1\cdot \text{Ai}\left(-\epsilon_{y}\right)+c_2\cdot \text{Bi}\left(-\epsilon_{y}\right),\\
    0&=c_1\cdot \text{Ai}\left(\frac{L}{R}-\epsilon_{y}\right)+c_2\cdot \text{Bi}\left(\frac{L}{R}-\epsilon_{y}\right).\label{eq:boundary_sol2}
\end{align}
Since the Airy-functions are transcendental functions there is no closed form solution in general. Because of this, we are now considering two different limiting cases, namely the first where ${\frac{L}{R}\gg \epsilon_y}$ and the second, where $\frac{L}{R}-\epsilon_{y}<-1$. 

Considering the first case, where $\frac{L}{R}\gg \epsilon_y$, simply corresponds to the system, where the energy of the particle in $y-$direction is far beyond the potential energy $mgL$ at the upper boundary. Due to that, one can apply the asymptotic properties of the Airy-function of second kind, namely
\begin{equation}
    \lim_{z\rightarrow \infty} \text{Bi}(z)=\infty.
\end{equation}
This means for Eq. \eqref{eq:boundary_sol2} in order for the total wave function to behave well a the boundary that $c_2=0$. The associated boundary condition $Y(0)=0=\text{Ai}(-\epsilon_y)$ yields the quantized values for $\epsilon_y$ and therefore energy eigenvalues in this limit. These energy values can e.g. be determined using a WKB approximation as shown in \cite{Sakurai}. The result reads
\begin{equation}
    \epsilon_{y,k}=\left(\frac{3\pi}{2}\cdot (k-\frac{1}{4})\right)^{\frac{2}{3}}~~~k\in \mathbb{N}^+,
\end{equation}
and therefore
\begin{equation}\label{eq:qm_low_energy}
    E_{y,k}=\frac{\hbar^2}{2mR^2}\cdot \left(\frac{3\pi}{2}\cdot (k-\frac{1}{4})\right)^{\frac{2}{3}}.
\end{equation}
In order to distinguish the low energy from the high energy regime, we will use the quantum number $k$ for the first, and later on the quantum number $r$, for the other case.

In this limit the appropriate normalisation constant $c_1$ of the associated wave function can approximately be determined by the assumption $L\approx \infty$. For the interested reader the integrals involved in the integration are stated in the appendix \ref{sec:appendix}. The total wave function in the low energy limit thus reads
\begin{equation}
    Y_k(y)=\frac{\text{Ai}\left(\frac{y}{R}-\epsilon_{y,k}\right)}{\sqrt{R\cdot \text{Ai}'(-\epsilon_{y,k})^2}}.
\end{equation}

As a second (and far more interesting case) we consider the situation $\frac{L}{R}-\epsilon_{y,n}<-1$ which is equivalent to the statement that the energy of the particle is larger than the potential energy at the upper boundary. Classically those cases correspond to bounces at $y=0$ and $y=L$. Note that the restriction $<-1$ is chosen in order to keep the errors in the following approximations of $\text{Ai}$ and $\text{Bi}$ below $1\%$. 
Solutions to this set of boundary equations \eqref{eq:boundary_sol1} \& \eqref{eq:boundary_sol2} are obtained if
\begin{equation} \label{eq:bound_cond_det}
    \det \begin{pmatrix}
    \text{Ai}(-\epsilon_{y,n}) & \text{Bi}(-\epsilon_{y,n})\\
    \text{Ai}\left(\frac{L}{R}-\epsilon_{y,n}\right) & \text{Bi}\left(\frac{L}{R}-\epsilon_{y,n}\right)
    \end{pmatrix}=0,
\end{equation}
is satisfied. To solve the last equation we can approximate the Airy-functions in the considered energy regime via
\begin{align}
    \text{Ai}(x)&\approx \frac{\sin\left(\frac{2}{3}(-x)^{\frac{3}{2}}+\frac{\pi}{4}\right)}{\sqrt{\pi}(-x)^\frac{1}{4}},\\
    \text{Bi}(x)&\approx \frac{\cos\left(\frac{2}{3}(-x)^{\frac{3}{2}}+\frac{\pi}{4}\right)}{\sqrt{\pi}(-x)^\frac{1}{4}}.
\end{align}

After some straightforward calculations, including trigonometric identities, the condition \eqref{eq:bound_cond_det} can be written compact as

\begin{equation}
    \sin\left(\frac{2}{3}\left[\epsilon_{y}^\frac{3}{2}-(\epsilon_{y}-\frac{L}{R})^\frac{3}{2}\right]\right)=0.
\end{equation}
Clearly it is satisfied if
\begin{equation}
    \epsilon_{y}^\frac{3}{2}-(\epsilon_{y}-\frac{L}{R})^\frac{3}{2}=\frac{3r\pi}{2}~~~r\in \mathbb{N}^+.
\end{equation}

The solutions for the reduced energy eigenvalues $\epsilon_y$ depending on $r$ can not be written down in a closed form but due to our restriction $\frac{L}{R}-\epsilon_{y}<-1$ one can Taylor expand the equation up to second order in the variable $\frac{L}{R\epsilon_{y}}=w\approx 0$:

\begin{equation}
    \frac{3L}{2R}\cdot \sqrt{\epsilon_{y,r}}-\frac{3L^2}{8R^2\sqrt{\epsilon_{y,r}}}\approx \frac{3r\pi}{2}.
\end{equation}

Solving this equation and inserting the expression for $R$ results in the approximate quantized values for $\epsilon_y$
\begin{equation}
    \epsilon_{y,r}=\frac{r^2\pi^2R^2}{4L^2}\cdot \left[1+\sqrt{1+\frac{L^3}{R^3\pi^2r^2}}\right]^2,
\end{equation}
or in terms of the quantized energy eigenvalues
\begin{equation}
    E_{y,r}=r^2\cdot \frac{E_1}{4}\cdot \left[1+\sqrt{1+\frac{mgL}{E_1\cdot r^2}}\right]^2.
\end{equation}
Here $E_1=\frac{\pi^2\hbar^2}{2mL^2}$ is the ground state energy of a free particle of mass $m$ inside an infinite potential well of length $L$. In the large $r$ limit the asymptotic behavior of the energy equals that of a free particle in the infinite potential well with energy eigenvalues $E_{y,r}\approx E_1\cdot r^2$. This is the analogous quantum behavior as seen in the classical limit. Note that due to the restriction $\frac{L}{R}-\epsilon_{y,r}<-1$ the allowed values for $r$ depend on the chosen parameters $m$ and $L$ of our system. The complete wave function in $y-$direction thus reads
%\begin{widetext}
\begin{equation}\label{eq:Y_high}
    Y_r(y)=\frac{\text{Ai}\left(\frac{y}{R}-\epsilon_{y,r}\right)-\frac{\text{Ai}_-}{\text{Bi}_-}\cdot \text{Bi}\left(\frac{y}{R}-\epsilon_{y,r}\right)}{\sqrt{R\cdot\left[\left(\text{Ai}_-'-\frac{\text{Ai}_-}{\text{Bi}_-}\cdot \text{Bi}_-'\right)^2- \left(\text{Ai}_+'-\frac{\text{Ai}_-}{\text{Bi}_-}\cdot \text{Bi}_+'\right)^2\right]}}.
\end{equation}
%\end{widetext}
The correct normalization constant in terms of derivatives of Ariy-functions has been determined using the integral representations for the Airy-functions displayed in the appendix \ref{sec:appendix}.
In order to write the wave function in a compact form we introduced the short hand notation in the last equation, where the index minus refers to the value $-\epsilon_{y,r}$ in the argument of the function and similarly the index plus for $\frac{L}{R}-\epsilon_{y,r}$.

\section{Quantum Probability Density and Expected Positions}
\begin{figure}[htb]
    \centering
    \includegraphics[scale=0.65]{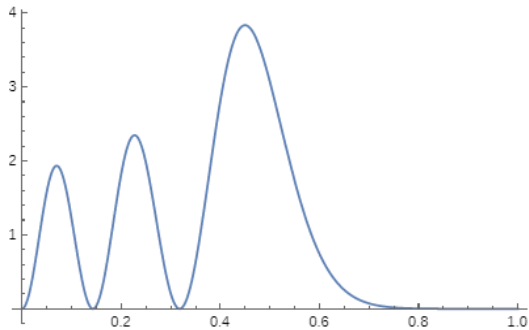}
    \caption{Probability density $|Y_3(y)|^2$ in the low energy limit with $R=0,1~L=1$ and $k=3$.}
    \label{fig:low_energy_k_3}
\end{figure}

\begin{figure}[htb]
    \centering
    \includegraphics[scale=0.65]{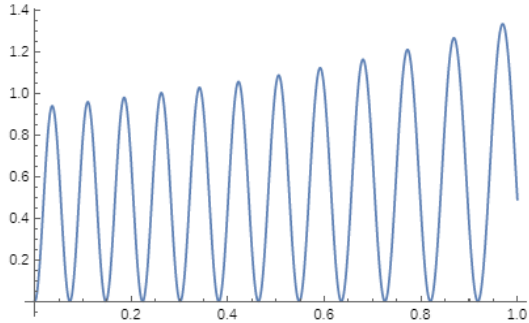}
    \caption{Probability density $|Y_{12}(y)|^2$ in the approximated high energy limit with $R=0,1,~L=1$ and $r=12$.}
    \label{fig:high_energy_k_10}
\end{figure}

In this final section we are considering the total probability densities for the quantum particle which by our separational ansatz is given by
\begin{equation}
    \varrho_{qm}(x,y)=|X_n(x)|^2\cdot |Y_{k/r}(y)|^2,
\end{equation}
where $k/r$ is understood to be either $k$ or $r$ depending on which energy regime we are considering in the $y-$direction. Note that the density separates in a $x-$ and $y-$depending part $\varrho_{qm}(x,y)=\varrho_{qm}(x)\cdot \varrho_{qm}(y)$. The $x-$depending part, as mentioned before, simply corresponds to a particle in the infinite potential well whose properties we already reviewed before. The interesting part, as in the classical study, again appears in the $y-$direction. In Figure \ref{fig:low_energy_k_3} the qualitative probability density $\varrho_{qm}(y)$ for $R=0,1$ as well as $L=1$ is displayed corresponding to the low energy limit where the particle (also classically) can not reach the upper boundary at $y=L=1$.
An analogous qualitative plot for the probability density in $y-$direction in the high energy case (approaching the same behavior as a free particle in the infinite potential well) is shown in Figure \ref{fig:high_energy_k_10}. Note that the fact that the wave function, and therefore the probability density, is not exactly zero at $y=L$ results from the minimal error related to the Taylor expansion for the energy eigenvalues.

Note that both graphs show qualitatively the same behavior as in the classical regime shown in Figure \ref{fig:class_prob_examples}. For a three-dimensional representation of the total probability density, consult the graphics displayed in appendix \ref{sec:appendix}.

We close this section with a calculation of the position expectation values and related uncertainties. The $x-$direction quantities can easily be derived (see e.g. \cite{Gupta}) and read in terms of the classical results Eqs. \eqref{eq:class_x_expectation} \& \eqref{eq:class_standard_x_deviation}
\begin{align}
    \langle x\rangle_{qm}&=\langle x\rangle_{cl},\\
    \Delta x_{qm}&=\Delta x_{cl}\cdot \sqrt{1-\frac{6}{\pi^2n^2}},
\end{align}
depending on the quantum number $n$.

For the $y-$component quantities we consider the low and high energy limit separately again. The results for the low energy (see Eq. \eqref{eq:qm_low_energy}) limit, basically corresponding to a particle without upper boundary, have e.g. been calculated in \cite{Gupta} and read
\begin{align}
    \langle y\rangle_{qm}&=\frac{2}{3}\cdot \frac{E_{y,k}}{mg},\\
    \Delta y_{qm}&=\frac{2\sqrt{5}}{15}\cdot \frac{E_{y,k}}{mg}.
\end{align}
As in the classical case, we see the identical linear dependence if we associate the quantity $E_{y,k}/mg$ with the theoretical reachable maximal height $h$.

We now turn to the high energy limit where the wave function is given by \eqref{eq:Y_high}. The expectation value $\langle y\rangle_{qm}$ can be calculated in terms of the variable $z$ introduced in Eq. \eqref{eq:variables} via
\begin{equation}
    \langle y\rangle_{qm}=R\epsilon_{y,r}+R^2\int_{-\epsilon_{y,r}}^{\frac{L}{R}-\epsilon_{y,r}}z\cdot |Y_r(z)|^2dz.
\end{equation}
Performing the calculation, including the integral identities stated in the appendix \ref{sec:appendix} and using the boundary conditions for the wave function, it turns out that many terms appearing in the calculation cancel each other out, resulting in the simple expression for the position expectation value
\begin{equation}
    \langle y\rangle_{qm}=\frac{2}{3}\cdot \frac{E_{y,r}}{mg}-\frac{L}{3}\cdot \text{Ji}_+.
\end{equation}
In the last equation, we defined the new function $\text{Ji}_+$ analogous to the classical caseby the expression
\begin{equation}\label{eq:Ji}
    \frac{\left(\text{Ai}_+'-\frac{\text{Ai}_-}{\text{Bi}_-}\cdot \text{Bi}_+'\right)^2}{\left(\text{Ai}_-'-\frac{\text{Ai}_-}{\text{Bi}_-}\cdot \text{Bi}_-'\right)^2-\left(\text{Ai}_+'-\frac{\text{Ai}_-}{\text{Bi}_-}\cdot \text{Bi}_+'\right)^2},
\end{equation}
in order to write the formula for the expectation value in a compact form. Note that $\text{Ji}_+$ basically is $R^2\cdot Y_r'\left(\frac{L}{R}-\epsilon_{y,r}\right)^2$.
As in the classical result (see Eq. \eqref{eq:y_exp_cl}), we also see that the expected position in the quantum world is bounded from above by $2E_{y,r}/3mg$ and approaches, for large enough energies, the classical limit.

Let us finally consider the uncertainty in $y-$direction. Due to the translational invariance of the mathematical definition of the uncertainty, it follows that it can be calculated in terms of the variable $z$ as
\begin{equation}
    \begin{split}
 \Delta y_{qm}^2=R^3\int_{-\epsilon_{y,r}}^{\frac{L}{R}-\epsilon_{y,r}}z^2\cdot |Y_r(z)|^2dz\\
    -R^4\left(\int_{-\epsilon_{y,r}}^{\frac{L}{R}-\epsilon_{y,r}}z\cdot |Y_r(z)|^2dz\right)^2.
    \end{split}
\end{equation}

Using the integral identities involving Ariy-functions displayed in appendix \ref{sec:appendix} the calculations fo $\Delta y_{qm}$ can be carried out analytically. Remarkably, due to the boundary conditions, many terms in the lengthy calculation cancel out each other. The quantum uncertainty in terms of $\text{Ji}_+$ defined in Eq. \eqref{eq:Ji} for the large energy limit thus takes the form
%\begin{widetext}
\begin{equation}
    \Delta y_{qm}=\frac{2\sqrt{5}}{15}\cdot \frac{E_{y,r}}{mg}\cdot \sqrt{1+\frac{2mgL}{E_{y,r}}\cdot \text{Ji}_+-\frac{7}{2}\cdot \left(\frac{mgL}{E_{y,r}}\cdot \text{Ji}_+\right)^2}.
\end{equation}
%\end{widetext}
Again, as for the expectation value, the similarity to the classical result Eq. \eqref{eq:y_uncert_cl} is evident.

\section{Conclusion}
The classical and quantum study of a particle inside a box subjected to a gravitational field provided interesting results. The conditions on classical periodic orbits, including the interesting geometric phenomena happening at the boundary, have been worked out. A detailed discussion on the probability densities in both classical and quantum regime were derived and considered in specific limiting cases. In particular, the structural similarity of the subsequently obtained formulas for position expectation values, as well as uncertainties, yield a deeper insight into the connection between classical obtained results compared to the quantum ones.  For future work, it would be interesting to consider the (quantum) dynamics of systems under the influence of a gravitational field, where either the box is tilted by an angle or the boundary domain is not a square but e.g. a triangle or a parabola (referring to some of the classical study already performed e.g. in \cite{Masa_Grav}). 

%\newpage
\section{Appendix} \label{sec:appendix}
In this first part of the appendix, some integral identities involving Airy-functions in terms of the variables $z$ defined in Eq. \eqref{eq:variables}, which are used to calculate the normalization constant as well as expectation values, are stated. All of these identities have been obtained using Wolfram Alpha.

%\begin{widetext}
%\begin{footnotesize}
\begin{equation}
  I_1(z):=\int \text{Ai}(z)^2=z\cdot \text{Ai}(z)^2-\text{Ai}'(z)^2
\end{equation}

\begin{equation}
  I_2(z):=\int \text{Bi}(z)^2=z\cdot \text{Bi}(z)^2-\text{Bi}'(z)^2
\end{equation} 

\begin{equation}
    I_3(z):=\int \text{Ai}(z)\cdot \text{Bi}(z)=z\cdot \text{Ai}(z)\cdot \text{Bi}(z)-\text{Ai}'(z)\cdot \text{Bi}'(z)
\end{equation}
  
\begin{equation}
  I_4(z):=\int z\cdot \text{Ai}(z)^2dz=\frac{1}{6}\left(2z^2\text{Ai}(z)^2-2z\text{Ai}'(z)^2+2\text{Ai}(z)\text{Ai}'(z)\right)  
\end{equation}  
  
  $$I_5(z):=\int z\cdot \text{Ai}(z)\cdot \text{Bi}(z)dz$$
\begin{equation}
   =\frac{1}{6}\left(2z^2\text{Ai}(z)\text{Bi}(z)+\text{Ai}(z)\text{Bi}'(z)+\text{Ai}'(z)\text{Bi}(z)-2z\text{Ai}'(z)\text{Bi}'(z)\right) 
\end{equation}  
  
  $$I_6(z):=\int z\cdot \text{Bi}(z)^2dz$$
  
  \begin{equation}
  =\frac{1}{6}\left(2z^2\text{Bi}(z)^2-2z\text{Bi}'(z)^2+2\text{Bi}(z)\text{Bi}'(z)\right)
  \end{equation}
  
  $$I_7(z):=\int z^2\cdot \text{Ai}(z)^2dz$$
  \begin{equation}
  =\frac{1}{5}\left(\left(z^3-1\right)\text{Ai}(z)^2-z^2\text{Ai}'(z)^2+2z\text{Ai}(z)\text{Ai}'(z)\right)
  \end{equation}

 $$ I_8(z):=\int z^2\cdot \text{Ai}(z)\cdot \text{Bi}(z)dz$$
 \begin{equation}
     =\frac{1}{5}\left(\text{Ai}(z)\left(\left(z^3-1\right)\text{Bi}(z)+z\text{Bi}'(z)\right)+z\text{Ai}'(z)(\text{Bi}(z)-z\text{Bi}'(z))\right)
 \end{equation}

  $$I_9(z):=\int z^2\cdot \text{Bi}(z)^2dz$$
  \begin{equation}
      =\frac{1}{5}\left(\left(z^3-1\right)\text{Bi}(z)^2-z^2\text{Bi}'(z)^2+2z\text{Bi}(z)\text{Bi}'(z)\right)
  \end{equation}

%\end{footnotesize}
%\end{widetext}

%\newpage
Find next some three-dimensional plots for the quantum probability density in the low and high energy limit.

\begin{figure}[!htb]
    \centering
    \includegraphics[scale=0.65]{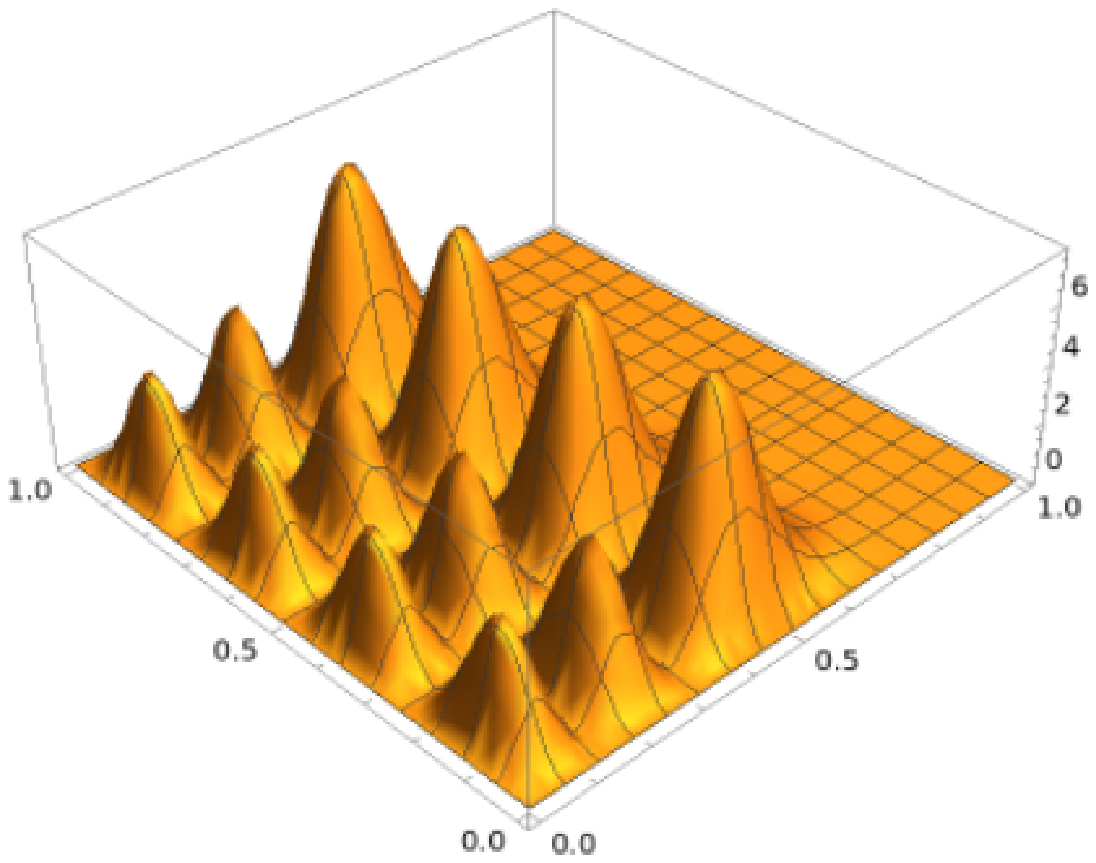}
    \caption{Graphical representation of the low energy probability density $\varrho_{qm}(x,y)=|X_4(x)|^2\cdot |Y_3(y)|^2$  for $R=0,1$ and $L=1$.}
   % \label{fig:my_label}
\end{figure}

\begin{figure}[!htb]
    \centering
    \includegraphics[scale=0.61]{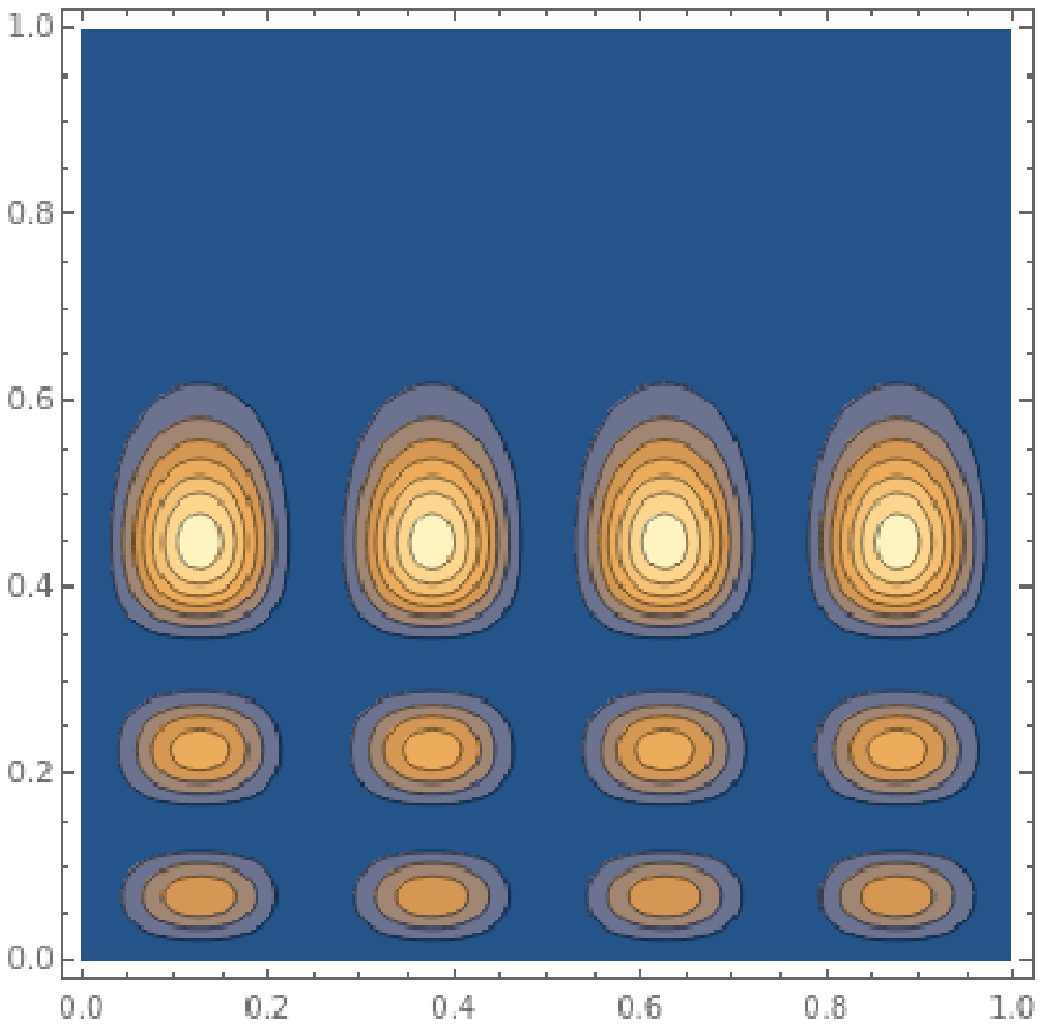}
    \caption{Contour plot of the low energy probability density $\varrho_{qm}(x,y)=|X_4(x)|^2\cdot |Y_3(y)|^2$  for $R=0,1$ and $L=1$.}
   % \label{fig:my_label}
\end{figure}

\begin{figure}[!htb]
    \centering
    \includegraphics[scale=0.61]{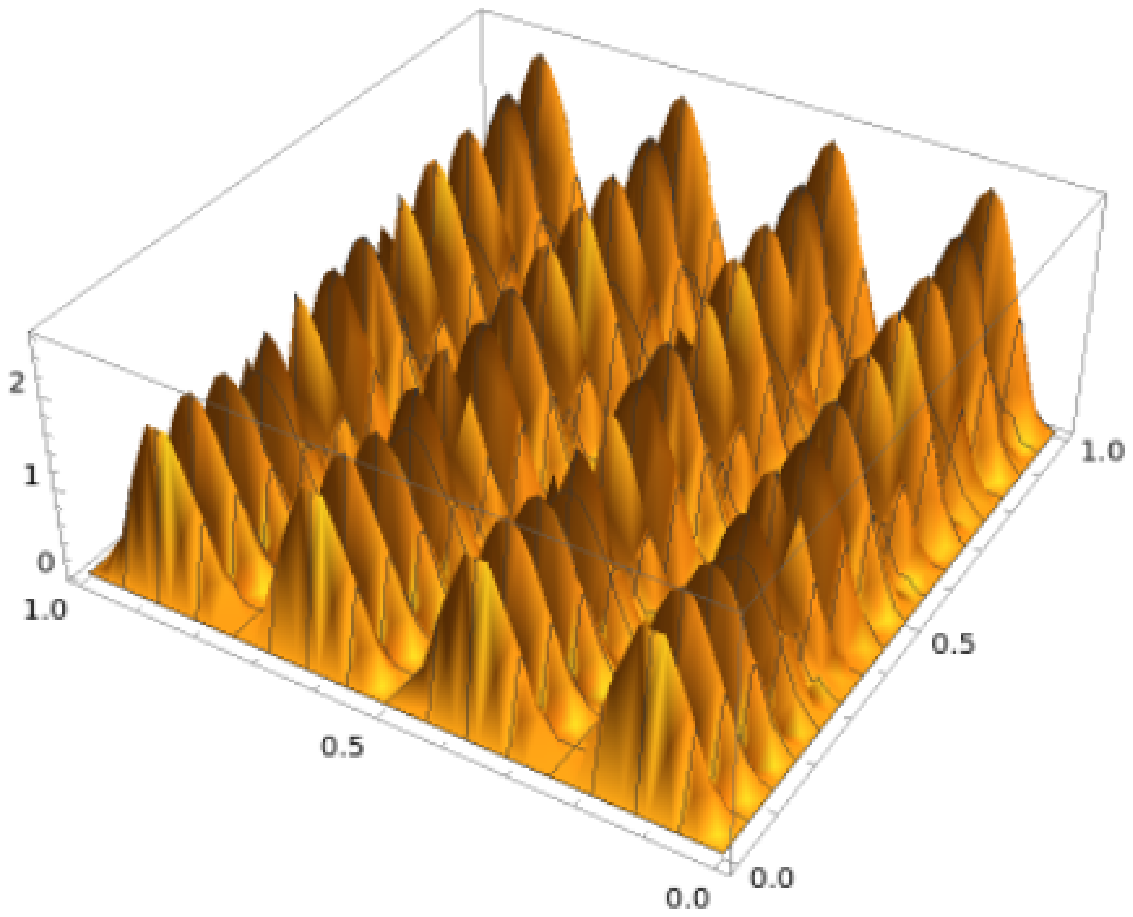}
    \caption{Graphical representation of the high energy probability density $\varrho_{qm}(x,y)=|X_4(x)|^2\cdot |Y_{12}(y)|^2$ for $R=0,1$ and $L=1$.}
   % \label{fig:my_label}
\end{figure}

\begin{figure}[!htb]
    \centering
    \includegraphics[scale=0.61]{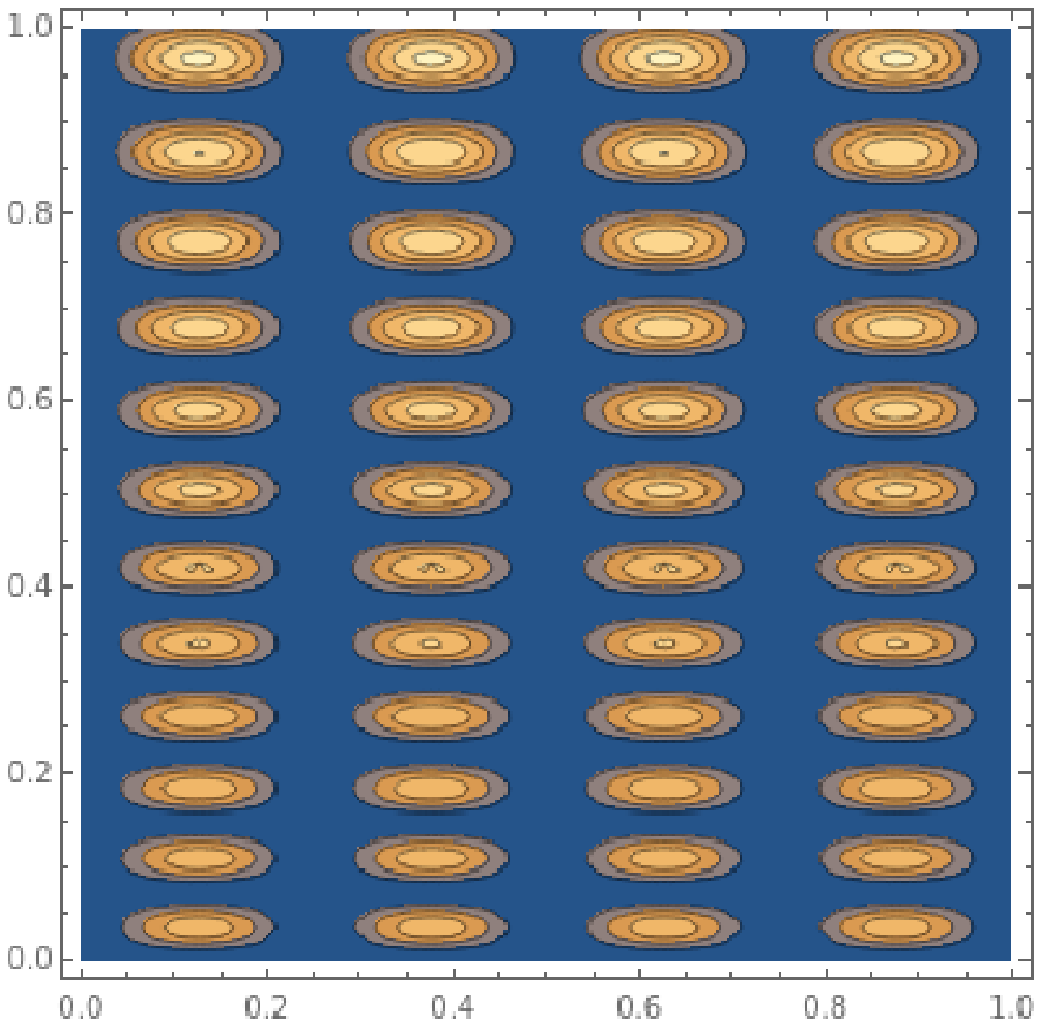}
    \caption{Contour plot of the high energy probability density $\varrho_{qm}(x,y)=|X_4(x)|^2\cdot |Y_{12}(y)|^2$ for $R=0,1$ and $L=1$.}
   % \label{fig:my_label}
\end{figure}

\newpage
\bibliographystyle{plain}

\begin{thebibliography}{9}
\bibitem{Grav_Bill_various}
D. R. da Costa, C. Dettmann \& E. D. Leonel, \href{https://doi.org/10.1016/j.cnsns.2014.08.030}{\textit{Circular, elliptic and oval billiards in a gravitational field}}, Communications in Nonlinear Science and Numerical Simulation, Volume 22, Issues 1–3, May 2015, Pages 731-746


\bibitem{Gupta}
A. Gupta, \href{http://home.iitk.ac.in/~gabhi/NIUSReport.pdf}{Classical and Quantum Probabilities}


\bibitem{Jaud1}
D. Jaud, \href{https://arxiv.org/pdf/1906.01911.pdf}{\textit{Playing a game of billiard with
Fibonacci
}}, arXiv:1906.01911v1 [math.DS], \url{DOI: 10.13140/RG.2.2.33902.46405/1},(2019)


\bibitem{Masa_Grav}
S. Masalovich, \href{https://arxiv.org/ftp/arxiv/papers/2007/2007.04730.pdf}{\textit{Billiards in a gravitational field: A particle bouncing on
a parabolic and right angle mirror}}, arXiv:2007.04730 [physics.optics], (2020) 


\bibitem{class_prob1}
D. A. B. Miller, \textit{Quantum Mechanics for Scientists and Engineers}, Cambridge: Cambridge University Press, (2007).


\bibitem{Circular_Well}
R. W. Robinett, \href{https://aapt.scitation.org/doi/10.1119/1.18188}{\textit{Visualizing the solutions for the circular infinite well in quantum and classical mechanics}}, American Journal of Physics 64, 440 (1996)


\bibitem{class_prob2}
R. W. Robinett, \href{https://doi.org/10.1119/1.17807}{\textit{Quantum and classical probability distributions for position and momentum}}, American Journal of Physics 63, 823 (1995)


\bibitem{Rozikov}
U. A. Rozikov, \textit{An Introduction to Mathematical Billiards}, World Scientific Publishing, (2019)


\bibitem{Sakurai}
J. J. Sakurai, \textit{Mordern Quantum Mechanics}, Addison-Wesley Publishing Company, (1994)

\bibitem{Tabach}
S. Tabachnikov, \href{http://www.personal.psu.edu/sot2/books/billiardsgeometry.pdf}{\textit{Geometry and Billiards}}, American Mathematical Society, (2005)


\bibitem{Airy_Book}
 O. Vall\'{e}e \& M. Soares  ,\href{https://doi.org/10.1142/p709}{\textit{Airy Functions and Applications to Physics}}, World Scientific Publishing, 2nd Edition, (2010)
\end{thebibliography}

\end{document}